\documentclass[fleqn,10pt]{wlscirep}

\usepackage{epstopdf}
\usepackage{float}
\usepackage[caption=false]{subfig}
\usepackage{ulem}
\usepackage{soul}

\title{Raman scattering excitation spectroscopy of monolayer WS$_2$}

\author[1,*]{Maciej R. Molas}
\author[1]{Karol Nogajewski}
\author[1,2]{Marek Potemski}
\author[2,*]{Adam Babi\'nski}
\affil[1]{Laboratoire National des Champs Magn\'etiques Intenses, CNRS-UGA-UPS-INSA-EMFL, 25, avenue des Martyrs, 38042 Grenoble, France}
\affil[2]{Faculty of Physics, University of Warsaw, ul. Pasteura 5, 02-093 Warszawa, Poland}

\affil[*]{maciej.molas@gmail.com, adam.babinski@fuw.edu.pl}

\begin{abstract}
	Resonant Raman scattering is investigated in monolayer WS$_2$ at low temperature with the aid of an unconventional technique, $i.e.$, Raman scattering excitation (RSE) spectroscopy. The RSE spectrum is made up by sweeping the excitation energy, when the detection energy is fixed in resonance with excitonic transitions related to either neutral or charged excitons. We demonstrate that the shape of the RSE spectrum strongly depends on the selected detection energy. The resonance of outgoing light with the neutral exciton leads to an extremely rich RSE spectrum, which displays several Raman scattering features not reported so far, while no clear effect on the associated background photoluminescence is observed. Instead, when the outgoing photons resonate with the negatively charged exciton, a strong enhancement of the related emission occurs. Presented results show that the RSE spectroscopy can be a useful technique to study electron-phonon interactions in thin layers of transition metal dichalcogenides.

\end{abstract}

\begin{document}
	
	\flushbottom
	\maketitle
	%
	%
	\thispagestyle{empty}
	
	\label{sec:intro}
	Raman scattering spectroscopy is an acknowledged characterization tool of layered materials, such as, for example graphene~\cite{Zhao2015SR,kang2015,Ding2016,Dai2015,ZhangX2016}, boron nitride~\cite{reich,cai2016,maschio}, or semiconducting transition metal dichalcogenides (TMDs)~\cite{fan,golasa,zhang_raman,delCorro,grzeszczyk}. Since the beginning of recent boom in the physics and technology of graphene-related systems, it has been primarily used to determine the number of layers in ultimately thin films of various 2D crystals serving as building blocks of more complex structures\cite{Malard,gorbachev,leeMoS2}. Additional advantages of this technique come out when Raman scattering spectra are investigated as a function of the excitation energy and/or under resonant conditions when either the incoming or outgoing photons coincide in energy with optically active electronic transitions~\cite{fan,golasa,zhang_raman,delCorro}. Resonant Raman scattering offers supplementary information on TMD layers concerning essentially the coupling of particular phonons to electronic transitions of a specific symmetry~\cite{soubelet,delCorro,delCorro_nano,grzeszczyk}.
	
	Here we report on resonant Raman scattering study of monolayer WS$_2$. An extremely rich Raman scattering response is uncovered when using an unconventional spectroscopy scheme, referred to as Raman scattering excitation (RSE). This method relies on tracing the Raman scattering response when the detection energy of the outgoing photons is fixed and the laser energy is being swept. The shape of the RSE spectrum strongly depends on a selected detection energy, in our case related to either the negatively charged (X$^-$) or the neutral (X$^0$) exciton. The former resonance condition results in a strong enhancement of the X$^-$ emission due to cascade scattering by optical and acoustic phonons. In contrast, several Raman scattering processes, including double out-of-plane A'$_{1}$ zone-centre modes, are enhanced while being in resonance with the outgoing photons emitted at the energy of the X$^0$ exciton. The RSE spectroscopy is proposed as a convenient tool to investigate electron-phonon interactions in thin layers of TMDs.

	\section*{Experimental results}\label{sec:results}
	
	\begin{figure}[t]
		\centering
		\includegraphics[width=0.8\linewidth]{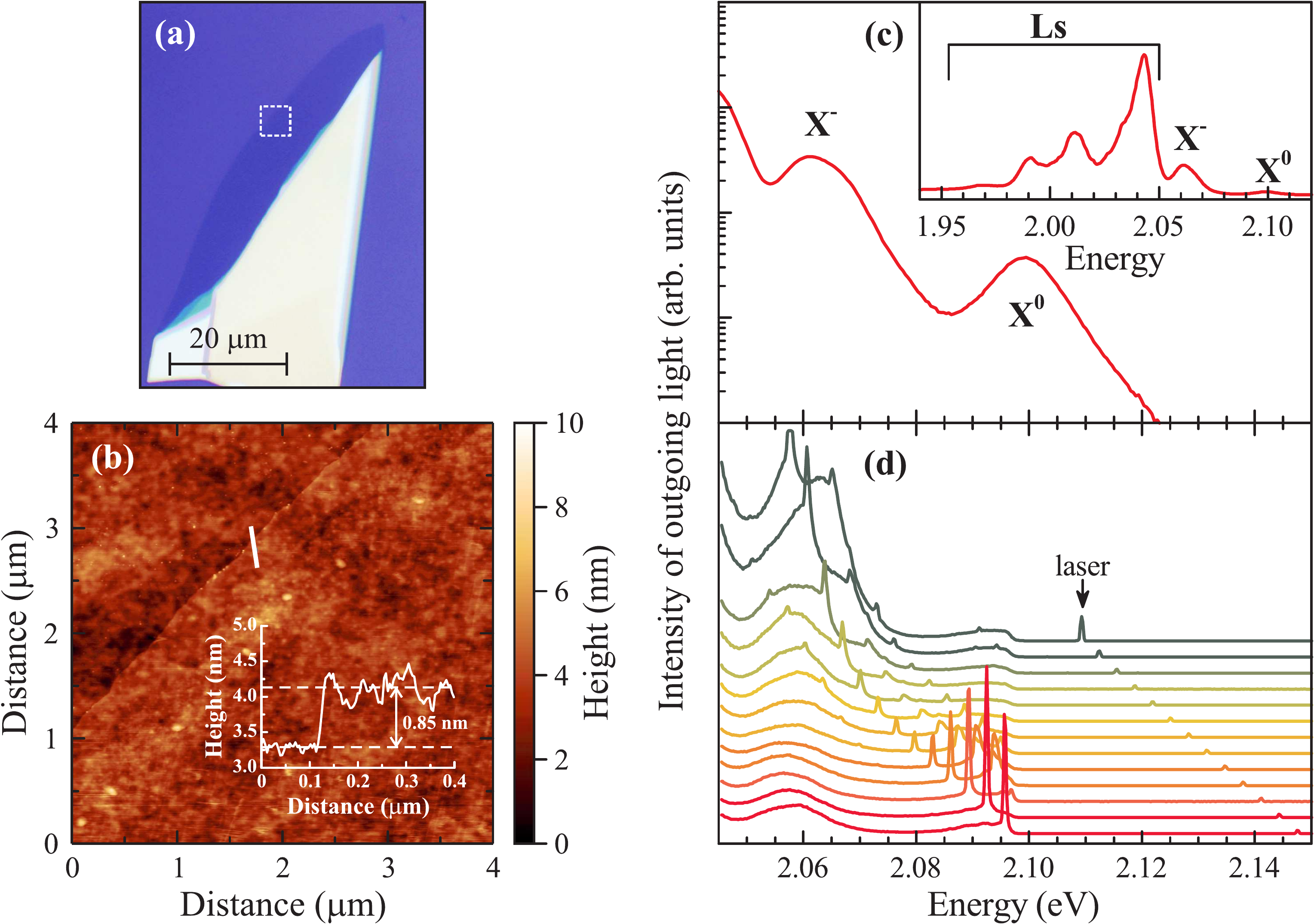}
		\caption{(\textbf{a}) Optical microscope image of the investigated WS$_2$ monolayer. (\textbf{b}) False-colour atomic-force-microscope image of an area enclosed in (\textbf{a}) with a dashed white box. Shown in the inset is a height profile measured along the white line crossing the edge of the monolayer. (\textbf{c}) Representative PL spectrum of the WS$_2$ monolayer from panel (\textbf{a}) related to the X$^-$ and X$^0$ emission, recorded at $T$=5~K with the use of non-resonant excitation at $\lambda=514.5\hspace{0.5mm}\mathrm{nm}$. The inset presents the PL spectrum in a broader energy range. Notice different intensity scales on the main graph (logarithmic scale) and the inset (linear scale). (\textbf{d}) Set of selected spectra of the outgoing light collected while tuning the excitation energy from $\sim$2.11~eV to $\sim$2.15~eV.}
		\label{fig:fig_1}
	\end{figure}
	
	Figure~\ref{fig:fig_1}(a) presents an optical microscope image of the investigated WS$_2$ monolayer supported by a Si/(300 nm)SiO$_2$ substrate. The topography of the flake is shown in Fig.~\ref{fig:fig_1}(b) with a false-colour atomic-force-microscope (AFM) image of an area enclosed in panel (a) with a dashed white box. The inset to panel (b) demonstrates a height profile measured along the white line crossing the edge of the monolayer. The thickness of the monolayer one can extract from that profile (0.85 nm) is slightly larger than the value of 0.62 nm (Ref.~\citenum{schutte}) based on the separation between adjacent layers in bulk WS$_2$ crystal. As reported also for other TMD monolayers deposited on Si/SiO$_2$ substrates, the difference may result from a non-zero equilibrium distance between the bottom surface of the monolayer and the top surface of the SiO$_2$ layer as well as from a thin layer of water captured between the monolayer and the substrate.	A representative photoluminescence (PL) spectrum of the monolayer excited at non-resonant energy  ($\lambda$=514.5~nm) is shown in Fig.~\ref{fig:fig_1}(c). It is composed of several emission lines (see the inset to Fig.~\ref{fig:fig_1}(c)), which are attributed to the neutral (X$^0$), negatively charged (X$^-$), and localized (Ls) excitons related to the maximum of the valence band and the higher-lying energy level in the conduction band at the K$^\pm$ points of the WS$_2$ monolayer's Brillouin zone (BZ)~\cite{mitioglu,ye,scrace,plechinger,shang,plechinger_nano,bala,molas,koperski}. Subsequent (selected) spectra obtained with several excitation energies are displayed in Fig.~\ref{fig:fig_1}(d). The spectral window presented covers the energy range that corresponds to the emission of light due to recombination of neutral and charged excitons. It can be seen that the lineshape of the spectra critically depends on the excitation energy. In particular, there are narrow lines superimposed on broad peaks due to recombination of the excitons. The lines follow the laser excitation energy which points out to Raman scattering as their origin. Moreover, the emission related to the negatively charged exciton is highly enhanced and broadened when its energy coincides with the energy of light scattered by processes identified later in the text.

	\begin{center}
		\begin{figure}[t]
			\centering
			\includegraphics[width=0.5\linewidth]{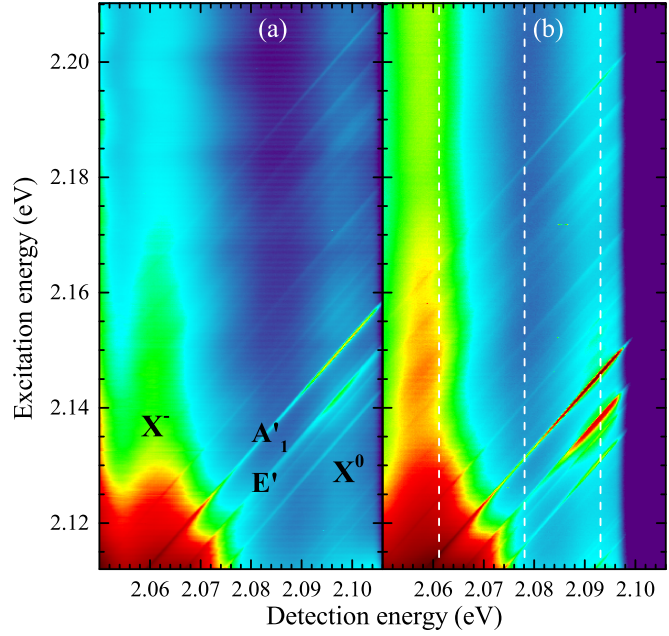}%
			\caption{Optical response of monolayer WS$_2$ at $T$=5~K plotted as a function of excitation energy. The excitation power was equal to (a) 10~$\mu$W and (b) 50~$\mu$W. The white dashed vertical lines denote the "cross-sections" of the map taken at (from the lowest energy) $E$=2.061~eV (X$^-$ emission), $E$=2.072~eV, and $E$=2.093~eV (X$^0$ emission) and presented in Fig.~\ref{fig:fig_3}}
			\label{fig:fig_2}
		\end{figure}
	\end{center}

\begin{figure}[t]
		\centering
		\includegraphics[width=0.5\linewidth]{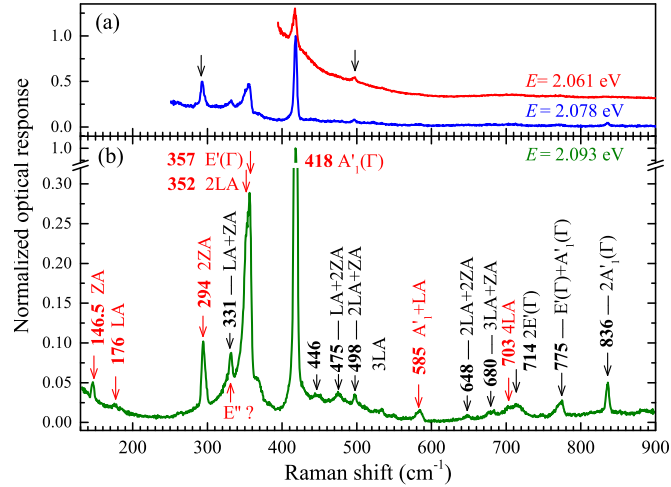}
		\caption{Low-temperature ($T$=5~K) Raman scattering excitation spectra of monolayer WS$_2$ detected at (a) $E$=2.061~eV and $E$=2.078~eV; (b) $E$=2.093~eV. The horizontal scale represents the Stokes shift between the excitation energy and the detection energy. The spectra are normalized to the most intense peaks. The assigned peaks correspond to phonons from the M point of the Brillouin zone unless otherwise stated. Features reported in earlier study are marked with red arrows.}
		\label{fig:fig_3}
	\end{figure}

	Complete sets of the collected data are presented in Fig.~\ref{fig:fig_2} in the form of color-coded maps showing the intensity of the optical response as a function of excitation energy. Panels (a) and (b) of the Figure present the results obtained with a lower ($\sim10~\mu W$) and higher ($\sim50~\mu W$) excitation power, respectively. Three energy regions of the optical response can be clearly distinguished in Fig.~\ref{fig:fig_2}: (i) the lowest-energy range ($<$2.075~eV) in which the emission ascribed to the X$^-$ exciton can be observed, (ii) an intermediate energy range (2.075~eV$\ldots$2.09~eV) in which a rather weak optical response can be seen, and (iii) the highest energy range ($>$2.09~eV) in which the emission due to the X$^0$ exciton becomes apparent. In order to identify the processes responsible for the Raman scattering in the three energy regions mentioned above, we have analyzed the optical response of the WS$_2$ monolayer measured as a function of excitation energy and detected at three selected energies: 2.061~eV, 2.078~eV, and 2.093~eV (see the white vertical dashed lines in Fig.~\ref{fig:fig_2}(b)). The resulting RSE spectra, displayed as a function of the Stokes shift, measured at 2.061~eV (in resonance with the negatively charged exciton), 2.078~eV, and 2.093~eV (in resonance with the neutral exciton) are shown in Fig.~\ref{fig:fig_3}(a) and Fig.~\ref{fig:fig_3}(b), respectively. When interpreting the spectra, it is important to keep in mind that although they correspond to Raman scattering, they were not obtained at a constant excitation energy, as it is typical for the Raman scattering technique. As a result, the spectra strongly depend on the detection energy, which corresponds to different resonance conditions. Particularly, the richest spectrum can be observed while the scattered light is in resonance with the neutral exciton, which points out to the role of exciton-phonon interactions in the Raman scattering processes.
	
	Let us focus first on the highest-energy region (see Fig.~\ref{fig:fig_3}(b)). 
	There are three zone-center ($\Gamma$) Raman active modes in monolayer 2H-TMDs, 
	which belong to the A'$_1$, E', and E" representations~\cite{molina}. 
	Two of them (A'$_1$ and E') are usually observed in the back-scattering configuration, in which the presented results have been obtained. The corresponding Raman scattering peaks can be seen in our experiment at 418~cm$^{-1}$ and 357~cm$^{-1}$, respectively. As it is widely accepted, the broadening of the latter line results from a double 2LA(M) process~\cite{Berkdemir2013}, whose peak emerges at 352~cm$^{-1}$ in the spectrum. In order to be Raman-active, the E" process requires the electric field of the laser light to be oriented out of the structure's plane, which is not possible in the back-scattering configuration. However, although the process is in principle forbidden in the geometry of our experiment~\cite{verble1970}, its observation under resonant excitation conditions is often reported in TMDs~\cite{nam2015,lee2015,soubelet}. We therefore ascribe the peak at 331~cm$^{-1}$ to the E" process as no other phonon mode is expected near this frequency. It should be noted that the energy of the peak matches quite closely the maximum of the total density of phonon states related to the phonon dispersion near the M point of the BZ~\cite{zhang_raman,lee2015}. Other peak related to phonons from the border of the BZ, namely ZA(M), is observed at 146.5~cm$^{-1}$. Its attribution to out-of-plane acoustic phonons was previously proposed in Ref.~\citenum{shi2016}. A weak feature at 176~cm$^{-1}$ can be attributed to a longitudinal acoustic mode from the M point of the BZ, LA(M)~\cite{Berkdemir2013}. Analyzing the Raman features related to the border of the BZ one must consider a possible presence of disorder in the structure. The phonon localization induced by disorder relaxes the momentum conservation rule, which allows the involvement of a single mode from outside of the $\Gamma$ point~\cite{frey1999,golasa2013}. Low intensity of the LA(M) feature suggests a rather low impact of disorder, while much higher intensities of the ZA(M) and E"(M) features are more likely due to the resonance with the neutral exciton. Following the assignment of the peak at 146.5~cm$^{-1}$ to the ZA(M) process it is naturally to ascribe the peak of a relatively high intensity observed at 294~cm$^{-1}$ to the process involving double out-of-plane acoustic phonons from the edge of the BZ, 2ZA(M)~\cite{shi2016}. Other features which were previously observed in the Raman scattering spectrum of monolayer WS$_2$ are combined processes: A'$_1$(M) + LA(M) at 585~cm$^{-1}$ and 4LA(M) at 703~cm$^{-1}$~\cite{Berkdemir2013}. We ascribe the peak occurring at 714~cm$^{-1}$ to the double 2E'($\Gamma$) process. The assignment of the peaks at 146.5~cm$^{-1}$ and 176~cm$^{-1}$ correspondingly to the ZA(M) and LA(M) phonons permits to propose an identification of other combined acoustic processes at 475~cm$^{-1}$, 498~cm$^{-1}$, 648~cm$^{-1}$, and 680~cm$^{-1}$ as LA(M) + 2ZA(M), 2LA(M) + ZA(M), 2LA(M) + 2ZA(M), and 3LA(M) + ZA(M), respectively. Moreover, the low-energy shoulder of the feature due to the E" process is close to the expected energy of the combined LA(M) + ZA(M) processes (322.5~cm$^{-1}$). Finally, the peaks at 775~cm$^{-1}$ and 836~cm$^{-1}$ can be attributed to the sum E'($\Gamma$)+A'$_1$($\Gamma$) and the double 2A'$_1$($\Gamma$) processes. The apparent broadening of the former feature may also result from the contribution of the 4LA(M) process. Both these features have not yet been reported for monolayer WS$_2$. 
	
\begin{figure}[t]
		\centering
		\includegraphics[width=0.4\linewidth]{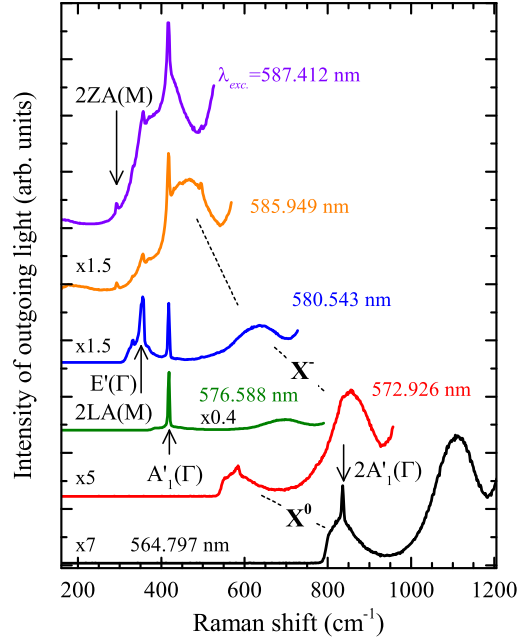}
		\caption{Optical response of monolayer WS$_2$ excited at $T$=5~K with laser light of different wavelengths.}
		\label{fig:fig_4}
	\end{figure}

	The RSE spectrum detected at 2.078~eV, in the intermediate energy range (see Fig.~\ref{fig:fig_3}(a)), is dominated by two Raman scattering modes: A'$_1$($\Gamma$) and 2LA(M)/E'($\Gamma$). A relatively high intensity of the peak due to the 2ZA(M) mode should be noted (see an arrow in Fig.~\ref{fig:fig_3}(a)). The optical response in the lowest-energy range (at 2.061~eV) which corresponds to the resonance of the scattered light with the charged exciton (see Fig.~\ref{fig:fig_3}(a)) shows an emission band of substantial intensity. As it can already be noticed in Fig.~\ref{fig:fig_2}, the emission due to the charged exciton is strongly enhanced by the resonance with two main Raman-active processes: A'$_1$($\Gamma$) and 2LA(M)/E'($\Gamma$). The overall emission is also shifted to higher energies while being in resonance with the light scattered by those two processes. When the excitation energy increases towards non-resonant conditions, the intensity saturates at substantially lower level. The difference between the resonant and non-resonant conditions may correspond to a fundamental change in the processes involved in the optical response. Out of the resonance only the PL emission is observed. In the resonance, the optical response is also due to cascade Raman scattering processes, which involve optical and acoustic phonons. In the latter case, the energy of the photoexcited carriers is lost in a series of scattering processes, while the intensity of the emission in the former process basically does not change with increasing the excitation energy. The difference between the two mechanisms responsible for the optical emission can be further appreciated by inspecting the spectra excited with laser light of particularly selected energies as shown in Fig.~\ref{fig:fig_4}. 
	
	\begin{center}
		\begin{figure*}[t]
			\subfloat{}%
			\centering
			\includegraphics[width=0.4\linewidth]{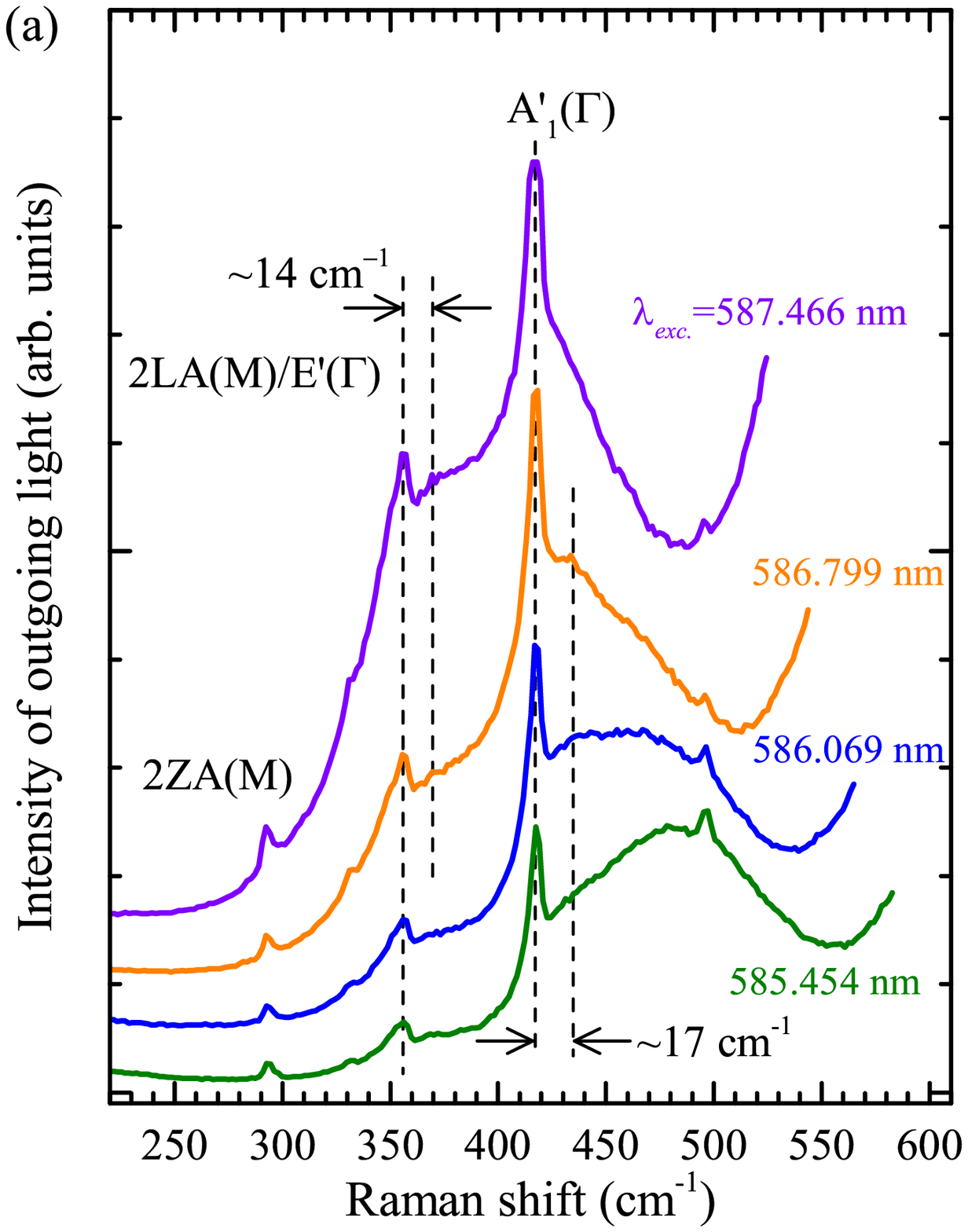}%
			\subfloat{}%
			\centering
			\includegraphics[width=0.4\linewidth]{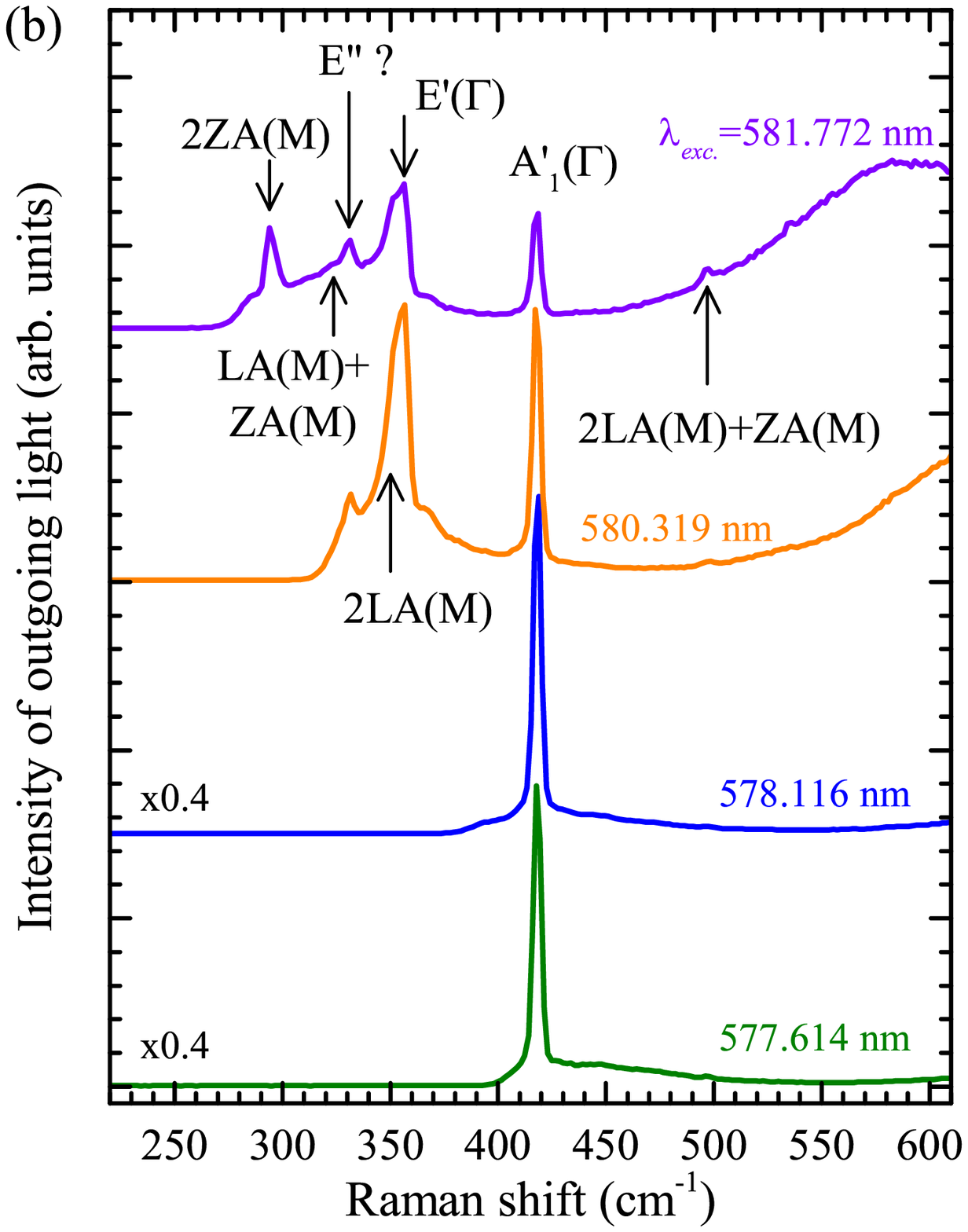}%
			\caption{Optical response of monolayer WS$_2$ at $T$=5~K excited with laser light of different wavelengths falling in the energy range of (a) charged and (b) neutral excitons.}
			\label{fig:fig_5}
		\end{figure*}
	\end{center}
	
	It is clearly seen in Fig.~\ref{fig:fig_4} that the intensity and lineshape of the optical spectrum related to the charged and neutral excitons strongly depend on the resonance conditions. The intensity of the optical response due to the charged exciton is strongly enhanced while being in outgoing resonance with main vibrational modes of the crystal lattice: 2LA(M)/E'($\Gamma$) or A'$_1$($\Gamma$) (see Fig.~\ref{fig:fig_5}(a)). The resonance of the light scattered by vibrational modes of the crystal with the neutral exciton results in the enhancement of the Raman scattering features with no strong effect on the background PL emission (see Fig.~\ref{fig:fig_5}(b)). Moreover, it can be seen in Fig.~\ref{fig:fig_5}(a), that the peak related to the 2LA(M)/E'($\Gamma$) process is accompanied by an additional structure whose maximum occurs $\sim$14~cm$^{-1}$ above. The structure is clearly visible while the outgoing resonance with the 2LA(M)/E'($\Gamma$) mode takes place at the energy corresponding to the maximum of the charged exciton peak and it becomes less pronounced outside that energy region. The lineshape of the spectrum around the resonance with the A'$_1$($\Gamma$) peak also suggests an additional contribution from higher-energy processes. A weak, broad feature at the energy higher by $\sim$17~cm$^{-1}$ than the energy of the A'$_1$($\Gamma$) peak can also be distinguished. Both high-energy structures must result from multiphonon processes, which involve principal optical phonons (2LA(M)/E'($\Gamma$) or A'$_1$($\Gamma$)) and additional acoustic phonons.
	
	Additional structures emerging on higher-energy slopes of the principal phonon modes in the investigated monolayer are related to multiphonon scattering by optical (LO) and acoustic phonons. The exciton-enhanced multiphonon Raman scattering can be explained in terms of a "cascade model". The process involves (1) an optical excitation of an exciton, (2) its relaxation with the emission of an optical phonon down to the vicinity of the band, (3) the subsequent emission of acoustic phonon(s), and (4) finally its radiative recombination~\cite{cardona}. The peak due to an additional acoustic process, which follows the emission of the optical phonon is broader than the LO phonon peak with the energy occurring at the largest value of the crystal momentum allowed by the exciton dispersion. The cascade scattering can be strongly enhanced by an outgoing resonance with excitonic complexes, as observed in several semiconductor systems~\cite{weisbuch}. A resonance with the recombination of a free electron and a hole localized on a carbon acceptor in GaAs (e, A$^0$) also leads to a similar effect~\cite{huang2001,huang2002}. Raman peaks related to the combined processes are dispersive, which reflects the exciton dispersion~\cite{yu1978}. No clear dispersion of the relevant peaks observed in our data most likely results from limited spectral resolution of the experimental set-up.  
	
	The resonance of the scattered light with the neutral exciton results in quite different spectra, as shown in Fig.~\ref{fig:fig_5}(b). The resonance induces a strong enhancement of several Raman peaks, clearly visible in Fig.~\ref{fig:fig_3}, but not of the background PL emission. This may point out to a different exciton-phonon interaction as compared with the charged exciton. In the case of the charged exciton, the strong optical response may be related to cascade Raman scattering involving both optical and acoustic phonons. The resonance with the neutral exciton gives rise mainly to the enhancement of Raman scattering by discrete modes. The enhanced modes, whose attribution has already been discussed, can clearly be seen in Fig.~\ref{fig:fig_3}(b). 
	
	Our results underline a complicated character of exciton-phonon interactions in thin TMD layers. This statement is even more valid in view of recent results reported by C.~M.~Chow~\cite{chow_arxiv} and co-workers, who showed virtually no effect of the Raman scattering on the emission due to the negatively charged exciton in monolayer MoSe$_2$ and a crucial effect of multiple LA(M) phonon emission on the neutral exciton. This might look surprising as both WS$_2$ and MoSe$_2$ share the same crystallographic structure. It is moreover known that critical differences between the influence of resonant excitation on the Raman scattering in different materials can exist~\cite{delCorro_nano}, which can be explained by theoretical calculations. The explanation presented in Ref.~\citenum{chow_arxiv} requires solid theoretical justification which is beyond the scope of our experimental work. We can, however, stress two points which may be important for the possible analysis. First is a crucial difference in the electronic structure of MoSe$_{2}$ and WS$_{2}$. Monolayer WS$_2$ is a darkish material, in which the energetically lowest transition is optically inactive~\cite{molas}. Monolayers of MoSe$_2$ are bright, which means that in their case the energetically lowest transition is optically active~\cite{molas}. Next, a closer inspection of our results shows that the resonantly enhanced emission due to the charged exciton is blue-shifted as compared to the emission excited out-of-resonance (see Fig.~\ref{fig:fig_2}(b)). Recently, it has been reported that the PL spectrum due to the negatively charged exciton in monolayer WS$_2$ is composed of two lines associated with two possible states of that exciton: intravalley (singlet) and intervalley (triplet) (for details see Ref.~\citenum{plechinger_comm}). In consequence, the observed blue shift may suggest that the resonance involves the intervalley (triplet) state of the charged exciton. It contrasts with monolayer MoSe$_2$, where the charged exciton is ascribed to the intervalley (singlet) state. This could explain the difference between the resonant-excitation effect on the charged exciton emission seen in our results and reported in Ref.~\citenum{chow_arxiv}. These facts may be of importance for the explanation of data and we do believe that they will trigger some interest in establishing their theoretical framework. 
	
	\section*{Conclusions}\label{sec:conclusions}
	
	We have presented a study of low-temperature optical emission from monolayer WS$_{2}$ excited resonantly in the energy range corresponding to the neutral and charged excitons. A clear difference between the Raman scattering excitation spectra detected at the energy of negatively charged and neutral excitons has been observed reflecting the differences in the electron-phonon interactions involved. The resonance of the emitted light with the negatively charged exciton results in the cascade scattering by the A'$_1$($\Gamma$), 2LA(M)/E'($\Gamma$) and acoustic phonons, which strongly enhances the related optical response of the system. The outgoing resonance with the neutral exciton leads to the enhancement of the Raman scattering intensity by several processes, including the double A'$_1$($\Gamma$) one. It has also been shown that the RSE spectroscopy employed in our experiment represents a sensitive tool to study electron-phonon interactions in thin films of TMD materials. 
	
	\section*{Methods}
	
	The WS$_2$ monolayer under investigation was prepared by mechanical exfoliation of a bulk crystal purchased from HQ Graphene. Initially, thin WS$_{2}$ flakes were exfoliated onto a polydimethylsiloxane (PDMS) stamp attached to a glass plate. The monolayers were then identified based on their optical contrast and cross-checked with the use of room-temperature Raman scattering and PL measurements. The highest quality ones of them were finally transferred to chemically cleaned and oxygen-plasma activated Si/(300 nm) SiO$_{2}$ substrates following a similar protocol as described in Ref.~\citenum{gomez}. Topography images of selected flakes were acquired using an NSV-VEECO-D3100 atomic force microscope operated in tapping mode under ambient conditions.

	Raman scattering measurements were carried out at low temperature ($T$=5~K) using a typical set-up for the PL and PL excitation experiments. The investigated sample was placed on a cold finger in a continuous flow cryostat mounted on x-y motorized positioners. The non-resonant PL measurements were carried out using 514.5 nm radiation from a continuous wave $Ar^+$ laser. To study the optical response of the system as a function of excitation energy, a dye laser based on Rhodamine 6G was used providing a tunable wavelength range extending from about 633 nm to almost 560 nm. The excitation light was focused by means of a 50x long-working distance objective (NA=0.50) producing a spot of about 1 $\mu$m diameter. The signal was collected via the same microscope objective, sent through a 0.5-m-long monochromator, and then detected by a charge-coupled device camera.

	\section*{Acknowledgements}
	
	The work has been supported by the European Research Council (MOMB project no. 320590), the EC Graphene Flagship project (no. 604391), the National Science Center (grant no. DEC-2013/10/M/ST3/00791), the Nanofab facility of the Institut N\'eel, CNRS UGA, and the ATOMOPTO project (TEAM programme of the Foundation for Polish Science co-financed by the EU within the ERDFund).
	
	\section*{Author contributions statement}
	
	M.R.M. carried out optical experiments and preliminary analysed the data. K.N. fabricated the samples under study and performed their AFM characterization. M.P. supervised the project and contributed to data analysis. A.B. performed the final data analysis. M.R.M. and A.B. wrote the paper with contributions from K.N. and M.P.
	
	\section*{Additional information}
	
	\textbf{Competing financial interests} The authors declare no competing financial interests. 
	
	\bibliography{biblio_raman}
		
\end{document}